\documentstyle[amstex,aps]{revtex}

\begin{document}
\draft
\title{Stability of narrow beams in bulk Kerr-type nonlinear media}
\author{B. A. Malomed}
\address{Department of Interdisciplinary studies \\
Faculty of Engineering \\
Tell Aviv University, Tel Aviv 69978, Israel \\
}
\author{K. Marinov\cite{byline} and D. I. Pushkarov}
\address{Institute of Solid State Physics \\
Bulgarian Academy of Sciences \\
BG-1784 Sofia, Bulgaria\\
}
\author{A. Shivarova}
\address{Faculty of Physics, Sofia University \\
BG-1164 Sofia, Bulgaria \\
}
\date{\today}
\maketitle

\begin{abstract}
We consider (2+1)-dimensional beams, whose transverse size may be comparable
to or smaller than the carrier wavelength, on the basis of an extended
version of the nonlinear Schr\"{o}dinger equation derived from the Maxwell`s
equations. As this equation is very cumbersome, we also study, in parallel
to it, its simplified version which keeps the most essential term: the term
which accounts for the {\it nonlinear diffraction}. The full equation
additionally includes terms generated by a deviation from the paraxial
approximation and by a longitudinal electric-field component in the beam.
Solitary-wave stationary solutions to both the full and simplified equations
are found, treating the terms which modify the nonlinear Schr\"{o}dinger
equation as perturbations. Within the framework of the perturbative
approach, a conserved power of the beam is obtained in an explicit form. It
is found that the nonlinear diffraction affects stationary beams much
stronger than nonparaxiality and longitudinal field. Stability of the beams
is directly tested by simulating the simplified equation, with initial
configurations taken as predicted by the perturbation theory. The
numerically generated solitary beams are always stable and never start to
collapse, although they display periodic internal vibrations, whose
amplitude decreases with the increase of the beam power.
\end{abstract}



\section{Introduction \newline
}

\label{sec:level1} Self-focusing of optical beams in nonlinear media \cite{1}
is a subject of intensive study since the critical nature of this phenomenon
(a collapse of the beam) has been identified in the first studies \cite
{2,3,4,5,6}. Different theoretical approaches have been developed for the
analysis of self-focusing, including the moment theory \cite{7}, variational
method (see, e.g., \cite{8} and references therein), paraxial-ray
approximation \cite{9}, ''adiabatic'' description \cite{10}, etc. With a
probability for a collapse, no stable self-trapped beams can exist in bulk
Kerr-media. Therefore, a considerable part of the studies in the field has
been directed towards finding mechanisms that limit the development of the
collapse and thus provide for the existence of stable self-trapped beams in
two transverse dimensions. It has been shown \cite{11} that the saturation
of the nonlinearity of the medium arrests the collapse and stable
two-dimensional self-trapped beams can indeed exist in such media \cite{8}.
The plasma generation by a self-focusing laser beam \cite{12,13,14} through
multiphoton and avalanche ionization, with its negative (exponential)
contribution to the refractive index of the media, can also stop the
collapse by leading to formation of filaments.

However, as it is pointed out in Ref. \cite{15}, the occurrence of the
self-focusing in different media (gases, liquids, solids) makes it
insufficient to identify a stabilizing effect for each separate case, and
calls to search for universal, i.e., medium-independent, mechanisms which
yield nonsingular behavior of the beams. In this connection, it is necessary
to stress on that the dynamics of narrow beams described by the full system
of the Maxwell`s equations may be drastically different from the predictions
of the usual parabolic nonlinear Schr\"{o}dinger (NLS) approximation valid
for broad beams (see, e.g., \cite{16} and references therein). In fact, the
Maxwell`s equations implicitly contain few different mechanisms that can
check the collapse of the narrow beams.

One of such mechanisms is nonparaxiality of the beam propagation, which
becomes important at the advanced stages of the self-focusing. It was
studied both numerically \cite{15,17,18,19} and analytically \cite{20},
concluding that the nonparaxiality replaces the catastrophic focusing with a
sequence of focusing-defocusing cycles. Beam filamentation due to the
nonparaxiality was demonstrated as well \cite{15}.

It was also shown that when the vector nature of the beam field is taken
into regard along with the nonparaxiality \cite{21}, the peak intensity
occuring in the self-focusing is almost ten times lower than that produced
by the nonparaxiality alone for the same input beam, suggesting that other
effects, which limit the self-focussing stronger than the nonparaxiality,
could exist. Recent analysis \cite{22,23} has demonstrated that the
efficiency of the vector model in limiting the narrowing of the beam in the
(1+1)-dimensional case is due not to the small longitudinal electric-field
component but rather to a scalar term which is present in the model. This
scalar term, accounting for the rate of the spatial variation of the
nonlinear polarization, combines effects of nonlinearity and diffraction. A
notation ''nonlinearly induced diffraction'' was introduced \cite{22,23} to
stress on this factor which restricts the beam narrowing predicted by the
usual NLS equation. In this connection it should be mentioned that a
fundamental limit for the soliton width has been found \cite{24,25,26} to
exist in the ''subwavelength'' range (beams with a transverse size
comparable with or smaller than the carrier wavelength) due to the terms
that come from Maxwell`s equations but are absent in the usual
(1+1)-dimensional NLS equation. Given the results in Ref. \cite{21} and
conclusions drawn in Refs. \cite{22,23,24,25,26}, the existence of a very
narrow stable self-trapped beam in an ideal bulk Kerr-medium is still an
open problem which needs to be addressed, which is the subject of the
present work.

The rest of the paper is organized as follows. In Section 2, we derive, from
the Maxwell`s equations, an equation to govern the evolution of narrow
(2+1)-dimensional beams in Kerr media. This is a very cumbersome
generalization of the NLS equation, with extra terms representing three
different effects: nonlinear diffraction, nonparaxiality, and the influence
of the small longitudinal component of the electric field. In view of the
very complex character of the equation, we also consider a simplified
version, that keeps only the nonlinear diffraction as a new effect. In
section 3, we develop a perturbation theory which treats all the new terms
in both equations as a perturbation, which is true in the case when the beam
is still relatively broad. Further analysis in section 4 shows that the
effect produced by the nonlinear diffraction is much stronger than those
generated by the nonparaxiality and longitudinal field, which justifies the
use of the simplified equation, at least for beams that are not extremely
narrow. Another noteworthy result of the perturbation theory is that,
although the terms which modify the NLS equation destroy the circular
symmetry of the beam, the deviation from the symmetry is, in fact, fairly
weak. In section 5, we show that the perturbative approach also allows one
to find an expression for the conserved power of the beam, which is not a
trivial issue in the present model. Lastly, in section 6 we display result
of direct numerical simulations of the beam propagation, starting from
initial configurations predicted by the perturbation theory. The direct
simulations are performed within the framework of the simplified equation
only. As a result, we conclude that the beams never show collapse and are
always stable, showing small internal vibrations, which are generated by a
deviation of the initial configuration from an exact solitary-beam shape.

\section{The beam propagation equation \newline
}

\label{sec:level2} We consider the propagation - along the $z$-axis - of a
light-beam, extended in the ($x$, $y$)-plane, in a bulk isotropic
Kerr-medium, characterized by the usual relation between the nonlinear
polarization $\vec{P}^{NL}$ and electric filed $\vec{E}$, $\vec{P}%
^{NL}=\varepsilon _{0}\chi \left( E_{x}^{2}+E_{z}^{2}\right) \vec{E}$, where 
$\chi $ is the $\chi _{xxxx}=\chi _{zzzz}$ component of the third-order
susceptibility tensor \cite{27}. The wave equation, 
\begin{equation}
\nabla ^{2}\vec{E}-\nabla (\nabla .\vec{E})=\frac{\varepsilon _{L}}{c^{2}}%
\frac{\partial ^{2}\vec{E}}{\partial t^{2}}+\mu _{0}\frac{\partial ^{2}\vec{P%
}_{NL}}{\partial t^{2}},  \label{eq1}
\end{equation}
combined with the relation 
\begin{equation}
\nabla .\vec{E}=-\frac{1}{\varepsilon _{0}\varepsilon _{L}}\nabla .\vec{P}%
_{NL},  \label{eq2}
\end{equation}
yield a scalar nonlinear wave equation in the following form:

\begin{eqnarray}
&&2i\beta \frac{\partial E}{\partial z}+\Delta _{\bot }E+\frac{3\omega
^{2}\chi }{4c^{2}}|E|^{2}E+\frac{3\chi }{4\varepsilon _{L}}\frac{\partial
^{2}\left( |E|^{2}E\right) }{\partial x^{2}}+\frac{\partial ^{2}E}{\partial
z^{2}}+\frac{\omega ^{2}\chi }{2c^{2}\beta ^{2}}\left| \frac{\partial E}{%
\partial x}\right| ^{2}E  \nonumber \\
&&-\frac{\omega ^{2}\chi }{4c^{2}\beta ^{2}}E^{\ast }\left( \frac{\partial E%
}{\partial x}\right) ^{2}+\frac{\chi }{\varepsilon _{L}}\frac{\partial }{%
\partial x}\left( \frac{E^{2}}{4}\frac{\partial E^{\ast }}{\partial x}%
\right) -\frac{\chi }{\varepsilon _{L}}\frac{\partial }{\partial x}\left( 
\frac{|E|^{2}}{2}\frac{\partial E}{\partial x}\right) =0.
\end{eqnarray}
In Eq.(3), $\omega $ is the frequency of the carrier wave, $E\equiv E_{x}$, $%
\Delta _{\bot }\equiv \left( \partial ^{2}/\partial x^{2}\right) +\left(
\partial ^{2}/\partial y^{2}\right) $ is the transversal Laplace operator, $%
\varepsilon _{L}$ and $\beta =\left( \omega /c\right) \varepsilon _{L}$ are
the linear permittivity and the linear propagation constant (wavenumber),
respectively. After rescaling $2\beta x\rightarrow x$, $2\beta y\rightarrow
y $, $2\beta z\rightarrow z$, $\gamma E\rightarrow E$, with $\gamma
^{2}=\left( 3\omega ^{2}\chi \right) /\left( 16\beta ^{2}c^{2}\right) $, and
some manipulation, Eq. (3) transforms into 
\begin{equation}
i\frac{\partial E}{\partial z}+\Delta _{\bot }E+|E|^{2}E+\frac{\partial ^{2}E%
}{\partial z^{2}}+4\left| \frac{\partial E}{\partial x}\right| ^{2}E-\frac{4%
}{3}\frac{\partial |E|^{2}}{\partial x}\frac{\partial E}{\partial x}+\frac{16%
}{3}\frac{\partial }{\partial x}\left( E\frac{\partial |E|^{2}}{\partial x}%
\right) =0.  \label{eq4}
\end{equation}
The modifications of the (2+1)-dimensional NLS equation involved in Eq.\ ($%
\ref{eq4}$) stem from nonparaxiality (the fourth term) and nonlinear
diffraction and longitudinal field component included in a combined manner
in the last three terms. It makes sense to consider Eq.\ ($\ref{eq4}$)
parallel to its ''model,'' simplified, version 
\begin{equation}
i\frac{\partial E}{\partial z}+\Delta _{\bot }E+|E|^{2}E+4\frac{\partial
^{2}\left( |E|^{2}E\right) }{\partial x^{2}}=0,  \label{eq5}
\end{equation}
where only the nonlinear diffraction [the last term in Eq.\ ($\ref{eq5}$)]
is taken into account to modify the NLS equation \cite{22,23}. The relevance
of approximating Eq.\ ($\ref{eq4}$) by Eq.\ ($\ref{eq5}$) will be verified 
{\it a posteriori}.

\section{Solitary-wave solutions and perturbation theory \newline
}

\label{sec:level3} Seeking for solitary-wave solutions to Eqs.\ ($\ref{eq4}$%
) and\ ($\ref{eq5}$) in the form $E(x,y,z)=F(x,y)$exp$(i\mu z)$, where $\mu $
is a nonlinear wavenumber shift of the solitary-wave solutions considered,
one obtains 
\begin{equation}
\left( 1+aF^{2}\right) \frac{\partial ^{2}F}{\partial x^{2}}+\frac{\partial
^{2}F}{\partial y^{2}}-kF+F^{3}+bF\left( \frac{\partial F}{\partial x}%
\right) ^{2}=0.  \label{eq6}
\end{equation}
In the case of Eq.\ ($\ref{eq5}$), $a=12$, $b=24$, and $k=\mu $, while Eq.\ (%
$\ref{eq4}$) corresponds to $a=32/3$, $b=68/3$ and $k=\mu +\mu ^{2}$.
Besides the linear diffraction in the second ($y$-) transverse direction,
Eq.\ ($\ref{eq6}$) exactly coincides with the one-dimensional equation
derived in Ref. \cite{22} for both the vector nonparaxial model and for the
scalar ''nonlinear diffraction'' model.

Generally speaking, Eq.\ ($\ref{eq6}$) must be solved numerically as it
stands. However, for the case of small gradients a perturbation scheme can
be developed, which simplifies the problem considerably. In fact, this
implies seeking for first corrections to the usual broad solitons, generated
by the new terms obtained from the Maxwell`s equations. To this end, $F(x,y)$
and $k$ are presented in the form $F(x,y)=F_{0}(x,y)+f(x,y)$, $%
k=k_{0}+\Delta k$, where $f$ and $\Delta k$ are the first order corrections,
so that the substitution of this into Eq.\ ($\ref{eq6}$) yields 
\begin{equation}
\frac{\partial ^{2}F_{0}}{\partial x^{2}}+\frac{\partial ^{2}F_{0}}{\partial
y^{2}}-k_{0}F_{0}+F_{0}^{3}=0,  \label{eq7}
\end{equation}
and 
\begin{equation}
\frac{\partial ^{2}f}{\partial x^{2}}+\frac{\partial ^{2}f}{\partial y^{2}}%
+\left( 3F_{0}^{2}-k_{0}\right) f=\Delta kF_{0}-aF_{0}^{2}\frac{\partial
^{2}F_{0}}{\partial x^{2}}-bF_{0}\left( \frac{\partial F_{0}}{\partial x}%
\right) ^{2}.  \label{eq8}
\end{equation}
In Eq.\ ($\ref{eq7}$), $k_{0}=\mu _{0}$ is the zeroth-order approximation
for the nonlinear wavenumber shift. The transformation to the polar
coordinates $x=r\cos \varphi ,\ y=r\sin \varphi $ transforms Eqs.\ ($\ref
{eq7}$) and\ ($\ref{eq8}$) into 
\begin{equation}
\frac{d^{2}F_{0}}{dr^{2}}+\frac{1}{r}\frac{dF_{0}}{dr}-k_{0}F_{0}+F_{0}^{3}=0
\label{eq9}
\end{equation}
and

\begin{eqnarray}
&&\frac{\partial ^{2}f}{\partial r^{2}}+\frac{1}{r}\frac{\partial f}{%
\partial r}+\frac{1}{r^{2}}\frac{\partial ^{2}f}{\partial \varphi ^{2}}%
+\left( 3F_{0}^{2}-k_{0}\right) f=\Delta kF_{0}-\frac{aF_{0}^{2}}{2}\left( 
\frac{d^{2}F_{0}}{dr^{2}}+\frac{1}{r}\frac{dF_{0}}{dr}\right)  \nonumber \\
&&-\frac{bF_{0}}{2}\left( \frac{dF_{0}}{dx}\right) ^{2}-\left[ \frac{%
aF_{0}^{2}}{2}\left( \frac{d^{2}F_{0}}{dr^{2}}-\frac{1}{r}\frac{dF_{0}}{dr}%
\right) +\frac{bF_{0}}{2}\left( \frac{dF_{0}}{dr}\right) ^{2}\right] \cos
2\varphi ,
\end{eqnarray}
respectively. Equation \ ($\ref{eq9}$), which is azimuthally symmetric,
gives rise to the classical Townes soliton \cite{5}, while Eq. (10)
generates a first-order correction to it. In Eq. (10) the variables can be
separated by means of the substitution $f(r,\varphi )=f_{0}(r)+f_{1}(r)\cos
2\varphi $, the resulting equations for $f_{0}$ and $f_{1}$ being 
\begin{equation}
\frac{d^{2}f_{0}}{dr^{2}}+\frac{1}{r}\frac{df_{0}}{dr}+\left(
3F_{0}^{2}-k_{0}\right) f_{0}=\Delta kF_{0}-\frac{aF_{0}^{2}}{2}\left( \frac{%
d^{2}F_{0}}{dr^{2}}+\frac{1}{r}\frac{dF_{0}}{dr}\right) -\frac{bF_{0}}{2}%
\left( \frac{dF_{0}}{dr}\right) ^{2}  \label{eq11}
\end{equation}
\begin{equation}
\frac{d^{2}f_{1}}{dr^{2}}+\frac{1}{r}\frac{df_{1}}{dr}+\left(
3F_{0}^{2}-k_{0}-\frac{4}{r^{2}}\right) f_{1}=-\frac{aF_{0}^{2}}{2}\left( 
\frac{d^{2}F_{0}}{dr^{2}}-\frac{1}{r}\frac{dF_{0}}{dr}\right) -\frac{bF_{0}}{%
2}\left( \frac{dF_{0}}{dr}\right) ^{2}.  \label{eq12}
\end{equation}
Thus, the first-order perturbative solution to Eq.\ ($\ref{eq6}$) amounts to
solving three ordinary differential equations, viz., Eqs.\ ($\ref{eq9}$), \ (%
$\ref{eq11}$) and \ ($\ref{eq12}$). This can be easily done by means of the
shooting method \cite{28}. We supplement the equations with obvious boundary
conditions that single out soliton solutions: 
\begin{gather*}
\frac{dF_{0}(r=0)}{dr}=\frac{df_{0}(r=0)}{dr}=\frac{df_{1}(r=0)}{dr}=0, \\
F_{0}(r=0)=A>0,f_{0}(r=0)=f_{1}(r=0)=0, \\
F_{0}(r\rightarrow \infty )\rightarrow 0,\,f_{0}\left( r\rightarrow \infty
\right) \rightarrow 0,\,f_{1}\left( r\rightarrow \infty \right) \rightarrow
0.
\end{gather*}

\section{Results of the perturbation theory \newline
}

\label{sec:level4} The solution of Eq. (9) is the same as that obtained in
Ref. \cite{5}. For this solution, $k_0\approx A^2/4.86$, and the power $P$
that it carries is $P=2\pi \int\limits_0^\infty rF_0^2dr\approx 11.70$,
which is the known "critical power" for the weak collapse in the
two-dimensional NLS equation \cite{5,7,8}.

Solutions to Eqs.\ ($\ref{eq11}$) and \ ($\ref{eq12}$) are displayed in
Figs. \ref{first_figure}(a) and \ref{first_figure}(b), respectively. The
solutions marked by "NLD" pertain to the case of $a=12$, $b=24$, which
corresponds to the simplified equation \ ($\ref{eq5}$), and those marked by
"vector" pertain to $a=32/3\approx 10.67$, $b=68/3\approx 22.67$, which
correspond to the full equation \ ($\ref{eq4}$). For comparison, the
zeroth-order solution (the Townes soliton) is shown in Fig. \ref
{first_figure}(c). It is obvious that the nonlinear diffraction is the main
factor which causes deviations from the solution of the NLS equation and the
effect of the longitudinal field component is quite small.

In Fig. \ref{second_figure}, results for the half-widths $R_{x}$, $R_{y}$ of
the beam (defined at half-maximum of its amplitude) are shown. It can again
be concluded that the nonlinear diffraction term is the main reason which
restricts the narrowing of the Townes soliton with the increase of its
on-axis intensity (denoted as ''standard'' in Fig. \ref{second_figure}). The
influence of the longitudinal field component shows up into a small
reduction of the effect of the nonlinear diffraction. It is also worth
noting that the half-width $R_{x}$ in the direction of the main
electric-filed component ($x$) is much more affected by the
nonlinear-diffraction term than that ($R_{y}$) in the other transverse
direction ($y$). The latter feature characterizes an effective $anisotropy$
induced by the new terms. Note that deviations from the circular symmetry of
a beam propagating according to an equation similar to Eq.\ ($\ref{eq4}$)
was reported in Refs.\cite{1,21}.

The comparison of the contributions to the nonlinear wavenumber shift $\mu $
stemming from nonparaxiality, nonlinear diffraction and longitudinal
electric field offers an additional way to estimate a relative strength of
effects generated by these factors. In the case when the simplified model \ (%
$\ref{eq5}$) is used, which keeps the nonlinear diffraction but neglects the
nonparaxiality and the influence of the longitudinal field, i.e., $a=12$ and 
$b=24$ in Eq.\ ($\ref{eq6}$), the result is $\mu _{NLID}\approx \frac{A^{2}}{%
4.86}+\frac{A^{4}}{1.74}$ up to the $A^{4}$-order. On the other hand, the
full model corresponding to Eq.\ ($\ref{eq4}$), i.e., with $a=32/3$ and $%
b=68/3$ in Eq.\ ($\ref{eq6}$), yields $\mu _{VN}\approx \frac{A^{2}}{4.86}+%
\frac{A^{4}}{1.74}\left( 1-\frac{1}{20.52}-\frac{1}{13.57}\right) $. The
first correction in the parenthesis is due to the vectorial structure of the
field, and the second one is an effect of the nonparaxiality. This result
demonstrates, that within the framework of the first-order perturbation
theory, the contribution to the nonlinear wavenumber shift stemming from the
nonlinear diffraction exceeds those originating from nonparaxiality and
longitudinal field component by approximately 13 and 20 times, respectively,
thus validating the use of Eq. \ ($\ref{eq5}$) instead of the full equation
\ ($\ref{eq4}$) for the study of the solitary waves.

\section{An approximate power-conservation law \newline
}

\label{sec:level5} According to the results of the perturbation theory for
the case of small gradients (Figs. \ref{first_figure} and \ref{second_figure}%
), the deviation of the shape of the beam from the circular symmetry is
small (although existing). Introducing the polar coordinates into Eq. \ ($%
\ref{eq5}$) and assuming weak dependence of the field envelop on the angular
variable $\varphi $, one obtains 
\begin{equation}
i\frac{\partial E}{\partial z}+\left( \frac{\partial ^{2}}{\partial r^{2}}+%
\frac{1}{r}\frac{\partial }{\partial r}\right) \left( E+2|E|^{2}E\right) +%
\frac{1}{r^{2}}\frac{\partial ^{2}E}{\partial \varphi ^{2}}+|E|^{2}E+2\cos
2\varphi \left( \frac{\partial ^{2}}{\partial r^{2}}-\frac{1}{r}\frac{%
\partial }{\partial r}\right) \left( |E|^{2}E\right) =0.  \label{eq13}
\end{equation}
Multiplying Eq.\ ($\ref{eq13})$ by $\left( E^{\ast }+|E|^{2}E^{\ast }\right) 
$, subtracting the complex conjugate and integrating over the transversal
plane while keeping the first order corrections only, a power conservation
law in the form 
\begin{equation}
\frac{dP}{dz}\equiv \frac{d}{dz}\int\limits_{0}^{2\pi
}\int\limits_{0}^{\infty }r\left( |E|^{2}+|E|^{4}\right) drd\varphi =0
\label{eq14}
\end{equation}
is obtained. Note that the second term in the integrand is due to the
nonlinear diffraction which comes to involve the power density associated
with the nonlinear polarization into the power conservation law.

In Fig. \ref{third_figure} the dependence of the power invariant, given by
Eq. (14), on the on-axis intensity of the beam is shown. For comparison, the
same dependence for the case of the standard NLS equation ($%
\int\limits_{0}^{2\pi }\int\limits_{0}^{\infty }r|E|^{2}drd\varphi $) is
also presented. It is evident that the solitary waves studied here exist for
power levels above the critical. The fact that the power is a growing
function of the on-axis intensity suggests stability for the solitary wave
solutions (see also the discussions in Ref. \cite{19} in the case of the
scalar nonparaxial model).

\section{Numerical simulations of the beam propagation \newline
}

\label{sec:level6} As Fig. \ref{third_figure} shows, the presence of
nonlinear diffraction increases the power necessary for self-trapping (i.e.,
the power at which the diffraction and nonlinearity balance each other).
This is demonstrated by the numerical simulations displayed in Fig. \ref
{fourth_figure}(a), where an initial configuration in the form of the Townes
soliton, see Eq.\ ($\ref{eq9})$, propagates according to the Eq.\ ($\ref{eq5}
$). We do not simulate the full equation \ ($\ref{eq4})$, which has an
extremely complex form, since the results presented above strongly suggest
that the simplified equation \ ($\ref{eq5}$) may describe essential features
of the beam dynamics quite adequately. Figure \ref{fourth_figure} shows that
in this case the beam diffracts away, while, as it is well known, the usual
NLS equation predicts collapse for the same initial conditions [see Fig. \ref
{fourth_figure}(c)]. In Fig. \ref{fourth_figure}(b), the approximate
invariant given by Eq.\ ($\ref{eq14})$ and the invariant of the NLS equation
in its standard form, $\int\limits_{-\infty }^{\infty }\int\limits_{-\infty
}^{\infty }|E|^{2}dxdy$, are shown vrs. the propagation distance for the
case when Eq.\ ($\ref{eq5}$) is solved. The fact that the NLS invariant
increases when the beam diffracts can be explained in the following way. As
it has been mentioned above, the full invariant, given by Eq.\ ($\ref{eq14})$%
, takes into account the power density associated with both the linear and
nonlinear polarization of the medium and is {\it conserved}, provided the
gradients remain small enough. When the beam diffracts, the contribution of
the nonlinear polarization decreases, transferring the power it carries to
the linear-polarization part, which therefore increases in accord with what
is seen in Fig. \ref{fourth_figure}(b). When the beam is self-focusing, the
computations show that the NLS invariant decreases, as one should expect.

In order to test directly the dynamical properties of the solitary waves in
the present model (first of all, their $stability$ against the collapse),
equation \ ($\ref{eq5})$ is solved numerically with initial conditions
produced by the perturbation scheme developed above [i.e., the solitary-wave
solutions generated by Eqs.\ ($\ref{eq9}$) -\ ($\ref{eq12})$ have been used
as initial configurations]. Figure \ref{fifth_figure} shows results for the
evolution of the beam intensity and of the approximate power invariant given
by Eq. \ ($\ref{eq14}$). For comparison, the usual NLS invariant is also
displayed. As is seen in Fig. \ref{fifth_figure}, the beams experience
periodic focusing and defocusing with a relatively small amplitude of the
corresponding internal vibrations, rather than the collapse that would occur
in the case of the NLS equation. The period of the oscillations rapidly
decreases with the increase of the input power.

The results presented in Figs. 5(a) and 5(b) are in a qualitative agreement
with those in Ref. \cite{18} where the non-catastrophycal propagation of
Gaussian beams within the scalar nonparaxial model has been obtained. The
soliton-like propagation studied later \cite{19} within the same model shows
the same type of oscillating behaviour. The influence of the deviations of
the initial conditions from the exact soliton-like solution has been also
checked. It seems that the vibrations obtained within the nonparaxial model
in Refs. \cite{18,19} and here, in a model which stresses on the nonlinearly
induced diffraction, are general behaviour which is a leftover of the former
collapse instability.

The oscillations observed here can be related to the fact that the initial
conditions, predicted by the perturbation theory, generate a
nearly-solitonic (but fairly stable) beam, rather than an exact stationary
soliton (cf. a similar situation in the multidimensional model with the
quadratic nonlinearity \cite{29}). However as the power of the beam
decreases and, accordingly, the beam radius increases, the accuracy of the
perturbation theory improves, hence the discrepancy between the exact
stationary self-trapped beams and those predicted by the perturbative
approach should decrease, i.e., lower-power beams should oscillate with a
smaller amplitude. This, however, is {\it not} the case. A closer look at
Fig. \ref{fifth_figure} shows that when the power is increased [from (a) to
(b)], the relative amplitude of the oscillations decreases slightly. To
check this trend, some results for the beam propagation at lower power
levels are presented in Fig. \ref{sixth_figure}. The three different curves
in this figure correspond to initial conditions in which, respectively, 6,
14, and 20 digits after the decimal point are kept to accurately implement
the prediction provided by the perturbation theory. The corresponding exact
values of the power are also given for each case. It is well seen from Fig.
6 that the evolution of the beam is quite sensitive to fine details of the
initial conditions. In order to achieve stationary propagation, the initial
conditions must be set with a very high accuracy, especially for lower power
beams. It turns out that the latter (wide) beams are more sensitive to small
deviations of their initial shape from exact solitary waves than the higher
power (narrow) beams. So, we infer that, quite naturally, the stabilizing
effect provided by the nonlinear diffraction gets enhanced with the decrease
of the beam size and the growth of its power.

\section{Conclusions \newline
}

\label{sec:level7}

Propagation of (2+1)-dimensional narrow beams in Kerr self-focusing media is
studied on the basis of an extended version of the NLS equation derived
directly from the Maxwell`s equations. As this equation is very cumbersome,
we also studied, in parallel to it, its simplified version which keeps the
most essential new term, viz., the one accounting for {\it nonlinear
diffraction}. The full equation contains, besides the nonlinear diffraction
term, terms generated by a deviation from the paraxial approximation and by
a longitudinal electric-field component in the beam. Solitary-wave solutions
to both the full and simplified equations are found, first, by means of a
perturbation theory which assumes that the new terms are still small,
provided that the beam is not too narrow. Within the framework of the
perturbative approach, small (but explicit) deviations of the beam from
circular symmetry are found, and the power-conservation law is derived in an
explicit form. It is shown that the nonlinear diffraction affects the
stationary beams much stronger than the nonparaxiality and longitudinal
field. Stability of the beams is tested by direct simulations of the
simplified equation, with initial configurations taken as predicted by the
perturbation theory. Numerically generated solitary beams are found to be
always stable and never sliding into collapse, although they display
periodic internal vibrations, whose amplitude decreases with the increase of
the beam power.

\acknowledgments

This work is supported by the National Fund for Scientific Research (project 
$n^0$ F-911) and by the Foundation for Scientific Research at Sofia
University (project $n^0$ 326/2000).

\begin{figure}[tbp]
\caption{Solutions of Eqs.\ ($\ref{eq11}$) and \ ($\ref{eq12}$),
respectively, in (a) and (b) with $a=32/3$, $b=68/3$ (notation ''vector'')
and $a=12$, $b=24$ (notation ''NLD'') as these quantities specify the
difference between Eqs.\ ($\ref{eq4}$) and \ ($\ref{eq5}$). In (c), the
solution of Eq.\ ($\ref{eq9}$) is presented together with the solutions $%
f_{0}$ and $f_{1}$ from (a) and (b) taken for the case of $a=12$, $b=24$. }
\label{first_figure}
\end{figure}
\begin{figure}[tbp]
\caption{Dependence of the half widths $R_{x}$ and $R_{y}$ on the on-axis
intensity $A^{2}$ [$A\equiv F_{0}(r=0)$] of the beam. The values of the
parameters are the same as denoted in Fig. 1(a). }
\label{second_figure}
\end{figure}
\begin{figure}[tbp]
\caption{Dependence of the power invariant $P$ on the on-axis intensity $%
A^{2}$ as given by Eq.\ ($\ref{eq14}$) [NLD model, Eq.\ ($\ref{eq5}$)] and
by the NLS equation. }
\label{third_figure}
\end{figure}
\begin{figure}[tbp]
\caption{On-axis intensity vrs. the propagation distance obtained as a
solution of Eq.\ ($\ref{eq5}$) in (a) and its corresponding power invariant
in (b); the power invariant corresponding to the NLS equation is given in
(b) in order to show its changes along the propagation distance. In (c), the
corresponding solutiuons of the NLS equation are shown. $A=0.11$ in the
initial condition. }
\label{fourth_figure}
\end{figure}
\begin{figure}[tbp]
\caption{Evolution of the normalized on-axis intensity [in (a) and (b)] as
solutions of Eq.\ ($\ref{eq5}$) and of the power invariant [in (c) and (d)]
according to Eq.\ ($\ref{eq14}$). For comparison the variations along the
propagation distance of the power invariant of the NLS equation are shown in
(c) and (d).}
\label{fifth_figure}
\end{figure}
\begin{figure}[tbp]
\caption{On-axis intensity vrs the propagation distance: check of
sensitivity with respect to the initial conditions.}
\label{sixth_figure}
\end{figure}

\end{document}